\documentclass[twocolumn]{aastex63}

\usepackage{lineno}
\usepackage{CJK}
\usepackage{amsbsy}
\usepackage{multirow}
\usepackage{bm}
\usepackage{ulem}

\received{XXX}
\revised{YYY}
\accepted{ZZZ}
\submitjournal{ApJ}

\shorttitle{Cosmic Web-Halo Connection Between Twin Universes}
\shortauthors{H.Z. Chen et al.}
\begin{document}
\begin{CJK*}{UTF8}{gbsn}

\title{Cosmic Web-Halo Connection Between Twin Universes}

\correspondingauthor{Xi Kang}
\email{kangxi@zju.edu.cn (ZJU)}

\author[0000-0001-8426-9493]{Hou-Zun Chen (陈厚尊)}
\affiliation{Purple Mountain Observatory, 10 Yuan Hua Road, Nanjing 210034, China}
\affiliation{School of Astronomy and Space Sciences, University of Science and Technology of China, Hefei 230026, China}

\author[0000-0002-5458-4254]{Xi Kang (康熙)}
\affiliation{Zhejiang University-Purple Mountain Observatory Joint Research Center for Astronomy, Zhejiang University, Hangzhou 310027, China}
\affiliation{Purple Mountain Observatory, 10 Yuan Hua Road, Nanjing 210034, China}

\author[0000-0003-2504-3835]{Peng Wang (王鹏)}
\affiliation{Leibniz-Institut f\"ur Astrophysik Potsdam, An der Sternwarte 16, D-14482 Potsdam, Germany}

\author{Noam I. Libeskind (李诺恩)}
\affiliation{Leibniz-Institut f\"ur Astrophysik Potsdam, An der Sternwarte 16, D-14482 Potsdam, Germany}
\affiliation{University of Lyon, UCB Lyon 1/CNRS/IN2P3, IPN Lyon, France}

\author[0000-0003-2341-9755]{Yu Luo (罗煜)}
\affiliation{Purple Mountain Observatory, 10 Yuan Hua Road, Nanjing 210034, China}
\affiliation{National Basic Science Data Center, Building No.2, 4, Zhongguancun South 4th Street, Haidian District, Beijing 100190, China}


\begin{abstract}

Both simulation and observational data have shown that the spin and shape of dark matter halos are correlated with their nearby large-scale environment. As structure formation on different scales is strongly coupled, it is trick to disentangle the formation of halo with the large-scale environment, making it difficult to infer which is the driving force for the correlation between halo spin/shape with the large-scale structure. In this paper, we use N-body simulation to produce twin Universes that share the same initial conditions on small scales but different on large scales. This is achieved by changing the random seeds for the phase of those $k$ modes smaller than a given scale in the initial conditions. In this way, we are able to disentangle the formation of halo and large-scale structure, making it possible to investigate how halo spin and shape correspond to the change of environment on large scales. We identify matching halo pairs in the twin simulations as those sharing the maximum number of identical particles within each other. Using these matched halo pairs, we study the cross match of halo spin and their correlation with the large-scale structure. It is found that when the large-scale environment changes (eigenvector) between the twin simulations, the halo spin has to rotate accordingly, although not significantly, to maintain the universal correlation seen in each simulation. Our results suggest that the large-scale structure is the main factor to drive the correlation between halo properties and their environment.

\end{abstract}

\keywords{Computational methods --- Astrostatistics --- Galaxy evolution --- Cold dark matter --- Large-scale structure of the universe}

\section{Introduction} \label{sec:1}

In the most widely accepted picture of structure formation, the initial seeds of gravitational instability are inherited from the adiabatic density fluctuations in the Big Bang, which can be well measured from the cosmic microwave background \citep{2014A&A...571A...1P}. After the recombination era, the competition between the gravitational force and cosmic expansion leads to the formation of dark matter halos at small scales in the regions where the linear density contrast surpasses the threshold \citep{1972ApJ...176....1G}. On large scales, both observations of galaxy surveys \citep[such as 2dF Galaxy Redshift Survey,][]{2003astro.ph..6581C} and Sloan Digital Sky Survey \citep{2000AJ....120.1579Y} and numerical simulations \citep[e.g.,][]{1985ApJ...292..371D} have revealed that the matter distribution can be characterized as a cosmic web (e.g. voids, walls, filaments, and clusters). As dark matter halos are residing in the cosmic web, one would expect that properties of dark matter halos are correlated with their surrounding cosmic environments or large-scale structures (hereafter LSS).  

The correlation between halo shape/spin and LSS has been extensively investigated in detail (see table 1 and table 2 in \citet{2014MNRAS.443.1090F}, also see table 1 in \citet{2018ApJ...866..138W}) in the last two decades in both observations and theoretical work.  On the theory side, many studies that used N-body and hydrodynamical simulations have reached almost converged conclusion that the major axes of halos/galaxies tend to be parallel with the slowest collapse direction of LSS, and the vector of angular momentum (spin) of halos/galaxies  is preferentially either aligned or perpendicular to the LSS, depending on the halo/galaxy mass \citep[e.g.,][]{2007A&A...474..315A, 2009ApJ...706..747Z, 2007MNRAS.375..489H, 2007MNRAS.381...41H, 2009ApJ...706..747Z, 2013ApJ...762...72T, 2017MNRAS.468L.123W, 2018MNRAS.473.1562W, 2018ApJ...866..138W, 2016IAUS..308..437C, 2020arXiv201201638L, 2014MNRAS.444.1453D, 2018MNRAS.481.4753C, 2019MNRAS.487.1607G, 2020MNRAS.493..362K}.

Along with the theoretical efforts, observational studies have focused on testing the results from simulations about the correlations between shape and spin of galaxy with the large scale structure. Using galaxy shape as a proxy of galaxy spin, several studies \citep[e.g.,][]{2009ApJ...706..747Z, 2013ApJ...775L..42T, 2013MNRAS.428.1827T, 2015ApJ...798...17Z, 2016MNRAS.457..695P, 2020MNRAS.491.2864W, 2020arXiv200304800M} claimed that galaxy spin of low-mass, spiral galaxies tend to align with the LSS, while high-mass, elliptical galaxies are preferentially perpendicular to the LSS, consistent with the simulation results. Some investigations \citep{2013ApJ...779..160Z} also found that red central galaxies show the strongest alignment signal between galaxies major axes with the host filaments or walls.

Generally speaking, the spin of a dark matter halo is the composition of the linear collapse of cosmological density field during the early stages, and the nonlinear collapse and mergers with other halos during the late stages. The former part can be well explained by the linear theory of the large scale tidal field, i.e., tidal torque theory \citep{1984ApJ...286...38W}. However, the later nonlinear part, which can only be studied with the help of N-body simulations, could be more dominant in the final angular momentum acquisition \citep{2002MNRAS.329..423M, 2002ApJ...581..799V}. For a long time, the correlations between halo properties and the LSS environment were usually explored in a statistical way using halo population as whole. If we focus on the merger history of every single halo, the individual differences can blur any conclusion that can be drawn. On the other hand, if we want to study the effect of the LSS direction or the environment, another halo sample with different merger histories has to be chosen from the simulation. It is thus difficult to study the effect of halo and environment separately. The degeneracy of LSS environment and nonlinear factors originating from small scales weaken our understanding of these halo-LSS conclusions mentioned above. Consequently, it's necessary to set up controlled N-body simulation in which we can somehow manipulate the large-scale environment of the halo and see how properties of given halos correspond to the changes of their LSS environments. In this work, we implement twin N-body simulations (explained in detail in Sec. \ref{sec:2}) to show how the spin (and major-axis) of matching halos evolves when the $\textbf{e}_3$ vector (i.e. the least compressed direction) changes its direction.

This paper is organized as follows. In Sec. \ref{sec:2} we present the method to produce twin universes using N-body simulation data and how to identify identical or matching dark matter halos.  In Sec. \ref{sec:3} we present the main results of the work. As a first step, we show the halo mass function and the correlations between halo spin/major-axis and $\textbf{e}_3$ vector of the LSS to check if the twin simulations are equal statistically. We then show how the spin/major axis of halo changes correspond to the change of the LSS environments. Lastly, the influence of LSS on halo shape parameters ($S$, $Q$ and $T$, see their definition in Sec. \ref{sec:3}) will be explored. Conclusions and discussions are presented in Sec. \ref{sec:4}.

\section{Methods and Simulations} \label{sec:2}

According to linear theory, the power spectrum of cosmic density perturbations and the two-point correlation function are Fourier transform pair, which can be expressed as follow:
\begin{equation}
\label{equ:equ1}
\xi\left(x\right)=\frac{1}{\left(2\pi\right)^3}\int P\left(k\right)\mathrm{e}^{i\textbf{k}\cdot\textbf{x}}\mathrm{d}^3\textbf{k}\;,
\end{equation}
where $P\left(k\right)=V_u\left\langle\vert\delta_{\textbf{k}}\vert^2\right\rangle$ is the power spectrum, $x=\vert\textbf{x}\vert$ is the distance, $k$ is the wave number, $V_u$ is the volume we consider. This relation often holds true for $V_u\rightarrow\infty$, or equivalently for $k\rightarrow0$. This indicates that the information of LSS in the late stages is totally contained in the small $k$ of the initial over-density field $\delta\left(k\right)$, just like fingerprints. 

In the most favoured frame for the generation of $\Lambda$CDM initial conditions, the power spectrum and the phase information are encapsulated as a Gaussian white noise field \citep{1996ApJ...460...59S}, which is particularly convenient to work with numerically. One can expect that the whole LSS pattern will change severely if we replace the random seeds for the phase of those $k$ mode smaller than a given scale, while the formation of dark matter halos on small scales will not be significantly affected. Besides, with the help of octree basis function \citep{2013MNRAS.434.2094J}, the LSS controlled initial conditions can also be generated in real-space \citep{2020arXiv200304321S}. This kind of technique paves a new way to manipulate the LSS environment of dark matter halos by a couple of N-body simulations.

In this work, we produce twin N-body simulations L500N1024, L500N1024a, L500N1024b, L500N1024c (hereafter N1024, N1024a, N1024b and N1024c). The letters ``a/b/c" here mark the three simulations for which the random seeds for small $k$ in the initial conditions have been changed, i.e., we replace the random seeds for the phase of $k$ mode such that:
\begin{equation}
\label{equ:equ2}
\vert\textbf{k}\vert<\frac{2\pi}{L_0}\;.
\end{equation}
Note that $L_{0}$ is an arbitrarily selected scale (discussed later) and should not be confused with the box-size of the simulation. In addition, we also run an independent simulation L500N1024r (hereafter N1024r) as a reference, for which the random seed for the phase of all $k$ modes is completely different from that used in the twin simulations. This reference simulation is only used to test the stochastic background of our method.

For the N-body simulations, the initial conditions are set at redshift $z=127$. At this initial time, the dark matter particles which later settle into a typical halo (with mass around $10^{13}M_{\odot}$ at $z=0$\textbf{)} occupied an initial volume about several cubic megaparsec. After some simple tests, we choose $L_0=20 h^{-1}\mathrm{Mpc}$. This scale is selected to produce significant changes of large-scale environments for most halos, but ensuring halo identities (more explicit definition in Sec. \ref{sec:2.1}) are not severely affected. Here we remind the reader that, in terms of statistical properties, all twin simulations are identical. Unless otherwise specified, all figures in this paper are plotted in a symmetric way, i.e., the conclusions do not change if we switch the N-body simulations. That is why we call them twin simulations. 

Our simulations were run using the P-Gadget-3 code, a TreePM code based on the publicly available code Gadget-2 \citep{2005MNRAS.364.1105S}, and they all contain $1024^3$ dark matter particles in a periodic box of $500h^{-1}\mathrm{Mpc}$. The cosmological parameters are chosen from the WMAP7 data \citep{2011ApJS..192...18K}, namely: $\Omega_{\Lambda}=0.7274$, $\Omega_m=0.2726$, $h=0.704$ and $\sigma_8=0.821$. The mass of single particle in our simulation is $8.8\times10^9\;h^{-1}\mathrm{M}_{\odot}$, and $136$ snapshots are stored from $z=127$ to $z=0$. We identify dark matter halos at each snapshot using the standard FoF (friend-of-friend) algorithm \citep{1985ApJ...292..371D} with a linking length that is 0.2 times the mean inter-particle separation. To ensure relatively robust results of halo spin and shape, and to avoid the spurious matches to the same halo purely by chance, the main halos containing at least 100 particles are chosen for further analysis. For each FoF halo, we identify subhalos using the SUBFIND algorithm \citep{2001MNRAS.328..726S} and construct merger trees for N1024 and N1024c, respectively.  Around $\sim568~000$ main halo candidates were identified from each simulation (see Tab. \ref{tab:1}).

\subsection{Match ratio and matching halos} \label{sec:2.1}

There is no unique way to match halos across N-body simulations. For example, \citet{2020arXiv200304321S} use the halo's $50$ most bound particles and a mass criterion to select matching halos. Here we simply define the match ratio between halos from the twin N-body simulations as Eq. \ref{equ:equ3}. Supposing main halo $A$ from N1024 contains $a$ dark matter particles, main halo $B$ from N1024a contains $b$ particles, and they have $c$ common particles according to the particle ids, then the match ratio is defined as:
\begin{equation}
\label{equ:equ3}
R\left(A,B\right)=\frac{c^2}{ab}\;.
\end{equation}
Generally speaking, when the random seeds for the phase of small $k$ are changed, a dark matter halo in N1024 may go through different evolutionary paths, such as being split into a few FoF groups, merge with other halos, or simply disappear. In other words main halo $A$ may be linked to multiple corresponding halos in N1024a. Each of them has a match ratio with main halo $A$ according to the Eq. \ref{equ:equ3} and the one with the largest ratio will be chosen as the matching halo of $A$ in N1024a (and vice versa). To ensure that there are enough matched halos for a statistical analysis, we set an arbitrary threshold to $R\left(A,B\right)\geq0.4$ for matching halo pairs, which roughly corresponds to each halo containing more than $65\%$ of its matching halo. After this selection, $157~748$ matching halo pairs were identified from N1024 and N1024a simulation at $z=0$. More information about the number of identified matching halo pairs are listed in Tab. \ref{tab:1} and Tab. \ref{tab:2}. Moreover, the number of main halos containing at least 100 particles at $z=2$ and $z=1$ is also listed.
\begin{table*}
\caption{The number of main halos (with more than 100 particles) found in each simulation at three different redshifts.}
\label{tab:1}
\centering
\begin{tabular}{c|c|c|c|c|c}
\hline\hline
& N1024 ($>$ 100) & N1024a ($>$ 100) & N1024b ($>$ 100) & N1024c ($>$ 100) & N1024r ($>$ 100) \\
\hline
$z=2$ & $309~586$ & $309~849$ & $309~269$ & $308~992$ & $310~298$  \\
$z=1$ & $491~356$ & $492~688$ & $492~913$ & $491~954$ & $492~282$  \\
$z=0$ & $568~247$ & $568~697$ & $569~947$ & $569~177$ & $567~560$  \\
\hline\hline
\end{tabular}
\end{table*}

\begin{table*}\footnotesize
\caption{The number of matching halos across the twin and the reference simulations at three different redshifts. In the twin simulation, there are more matching halo pairs. The matched halo with the reference simulation is very low, providing a noise background. As an example, we only list the result between N1024 and N1024r.}
\label{tab:2}
\centering
\begin{tabular}{c|c|c|c|c|c|c|c}
\hline\hline
& N1024/N1024a & N1024/N1024b & N1024/N1024c & N1024a/N1024b & N1024a/N1024c & N1024b/N1024c & N1024/N1024r\\
\hline
$z=2$ & $106~247$ & $105~597$ & $105~756$ & $106~833$ & $106~250$ & $106~794$ & $56$ \\
$z=1$ & $159~533$ & $159~258$ & $159~282$ & $160~819$ & $159~999$ & $160~608$ & $78$ \\
$z=0$ & $157~748$ & $156~646$ & $157~287$ & $158~444$ & $157~715$ & $158~560$ & $67$ \\
\hline\hline
\end{tabular}
\end{table*}

As mentioned in Sec. \ref{sec:1}, once the sample of matching halos between the twin N-body simulations is established, we can study the halo-LSS correlations. We employ the Hessian matrix method used in many previous articles \citep[e.g.,][]{2007MNRAS.375..489H, 2007MNRAS.381...41H, 2009ApJ...706..747Z, 2015ApJ...813....6K} to define the matrix as:
\begin{equation}
\label{equ:euq4}
H_{ij}=\frac{\partial^2\rho_s\left(\textbf{x}\right)}{\partial x_j\partial x_i}\;.
\end{equation}
This method is based on the smoothed density field $\rho_s\left(\textbf{x}\right)$ at the halo position  (based on a more accurate and improved algorithm by \cite{2020NewA...8001405W}), which can be given by the Cloud-in-Cell (CIC) technique \citep{1995pmtn.rept.....M}. The smoothing length $R_s$ is the only parameter of CIC, which can be regarded as the typical scale of halo LSS environment identified by Hessian matrix method. Many previous works \citep[e.g.,][but see also \citealt{2014MNRAS.441.1974L}]{2007A&A...474..315A, 2007MNRAS.381...41H, 2009ApJ...706..747Z, 2012MNRAS.427.3320C,2012MNRAS.425.2049H, 2013ApJ...762...72T,2013MNRAS.428.2489L} used a constant smoothing length. To determine which Rs value should be chosen, we test some LSS properties (e.g., environment, eigenvectors) of matching halos for three fixed Rs: $2.5,\;5,\;10h^{-1}\mathrm{Mpc} $. We find that the halo-LSS correlation, as well as other main conclusions we make below, become stronger as Rs decreases, which is a reasonable result according to our previous discussions. Consequently $R_s=2.5h^{-1}\mathrm{Mpc}$ is chosen hereafter. The three eigenvalues of the Hessian matrix are marked as $\lambda_1$, $\lambda_2$, $\lambda_3$, with corresponding eigenvectors $\textbf{e}_1$, $\textbf{e}_2$ and $\textbf{e}_3$. Eigenvalues $\lambda_i$ can be used to define the LSS environment of dark matter halos according to the number of positive eigenvalues \citep{1970A&A.....5...84Z, 2007MNRAS.375..489H, 2007MNRAS.381...41H, 2019RAA....19....6Z, 2018MNRAS.473.1195L}, i.e.,
\begin{itemize}
\item[1.] void:     $0<\lambda_1<\lambda_2<\lambda_3$
\item[2.] wall:     $\lambda_1<0<\lambda_2<\lambda_3$
\item[3.] filament: $\lambda_1<\lambda_2<0<\lambda_3$
\item[4.] cluster:  $\lambda_1<\lambda_2<\lambda_3<0$
\end{itemize}
and the eigenvectors $\textbf{e}_i$ stand for the three compressed directions of the smoothed density field. $\textbf{e}_3$ vector indicates the least compressed direction, which is a robust and universal definition of the LSS. In this work, we will focus on the alignment of halo spin and shape with the $\textbf{e}_3$ vector, i.e., $\cos\theta_3=\textbf{a}\cdot\textbf{e}_3$, where $\textbf{a}$ is the halo spin or major-axis vector.

\begin{figure*}[htp]
\plotone{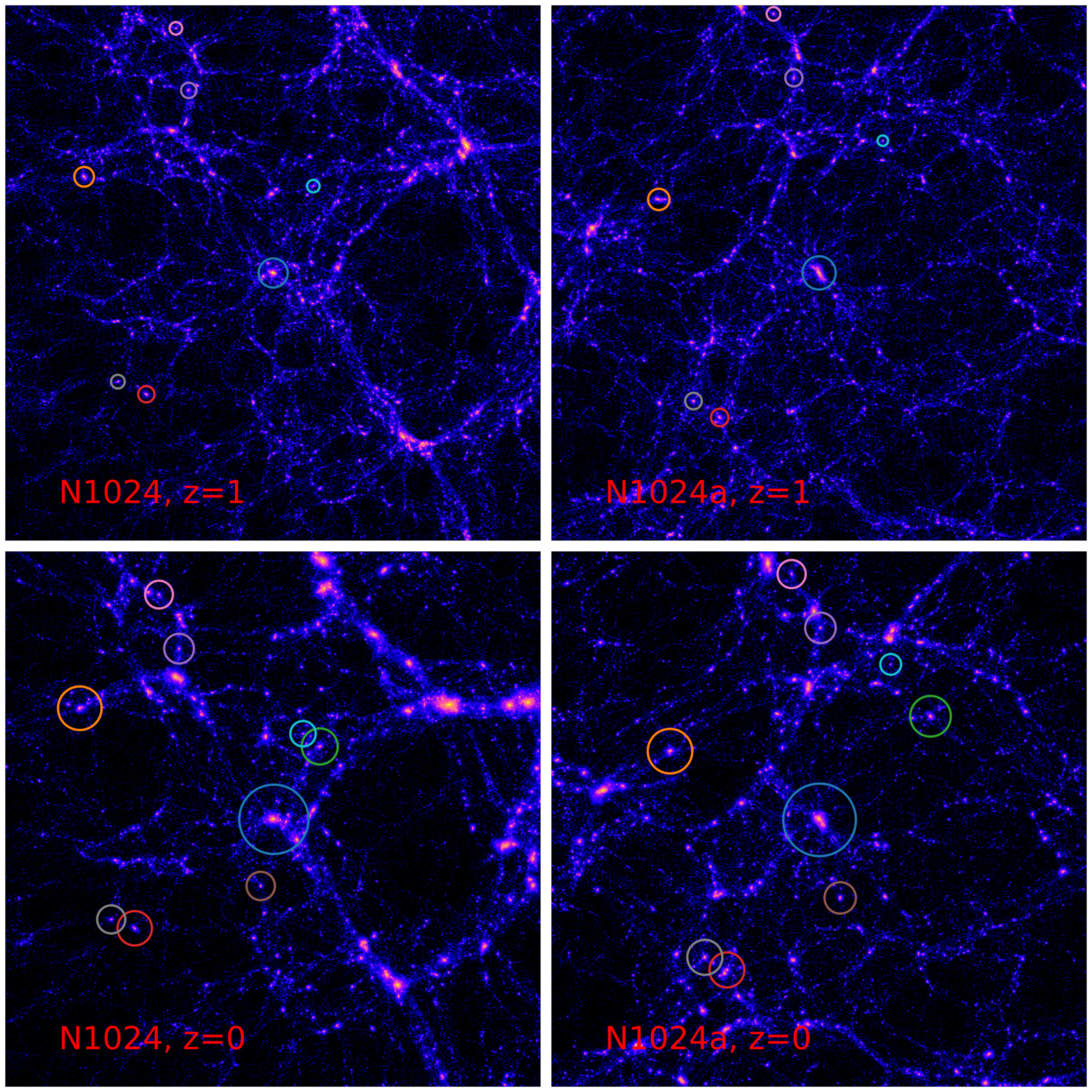}
\caption{Slices from twin simulations N1024 and N1024a, which are statistically equivalent.  A few matching halo pairs found at $z=0$ are shown in lower panels as colored circles with radius scaled to virial radius. Their corresponding progenitors at $z=1$ are shown in upper panels. For matched halo pairs, their LSS environments are visually different in the twin simulations. Note that the matching halos with green and brown circle disappeared at $z\sim1$, which means their progenitors are no longer matched at $z\sim1$.}
\label{fig:fig1}
\end{figure*}

Fig. \ref{fig:fig1} displays a $100\times100\times20h^{-1}\mathrm{Mpc}$ slice from the twin simulations N1024 and N1024a. It presents a global view of the twin simulations. Over two hundred matching halo pairs are identified in this displayed slice, but here we only mark nine matching halos with hollow circles, with size proportional to their virial radii. We marked their main progenitors in upper panels if and only if they are still matched at $z=1$. We can see that the LSS environment of these samples really changes tremendously after we replaced the random seeds for the small $k$ of the initial over-density field.

\section{Results} \label{sec:3}

Our main results, mostly regarding the influence of LSS environment transformation on the properties of dark matter halos, will be shown in this section. Before that, we show some statistical results, including the halo mass function and correlations between halo properties (spin and major-axis) and the LSS environment, to test whether the twin simulations are statistically equal. Then we will focus on the mass dependence of the match ratio $R$ in different LSS environment. Lastly, the influence of LSS on the total spin, major-axis, and shape parameters of the matching halo will be explored. Specifically, the shape parameters of dark matter halo, $S$, $Q$ and $T$, are defined by:
\begin{equation}
\label{equ:equ5}
S=\frac{c}{a}, \quad Q=\frac{b}{a}, \quad T=\frac{a^2-b^2}{a^2-c^2},
\end{equation}
where $a$ is the major-axis of modeled ellipsoid (by calculating the inertia tensor) of FoF halo, $b$ is the intermediate-axis, and $c$ is the minor-axis. The parameters $S$ and $Q$ give the ellipticity of dark matter halos, and $T$ quantifies the triaxiality \citep{1991ApJ...383..112F, 1992ApJ...399..405W}. Purely prolate halos have $T=1$ while purely oblate halos have $T=0$.

\subsection{Statistical tests} \label{sec:3.1}

\begin{figure}[htp]
\plotone{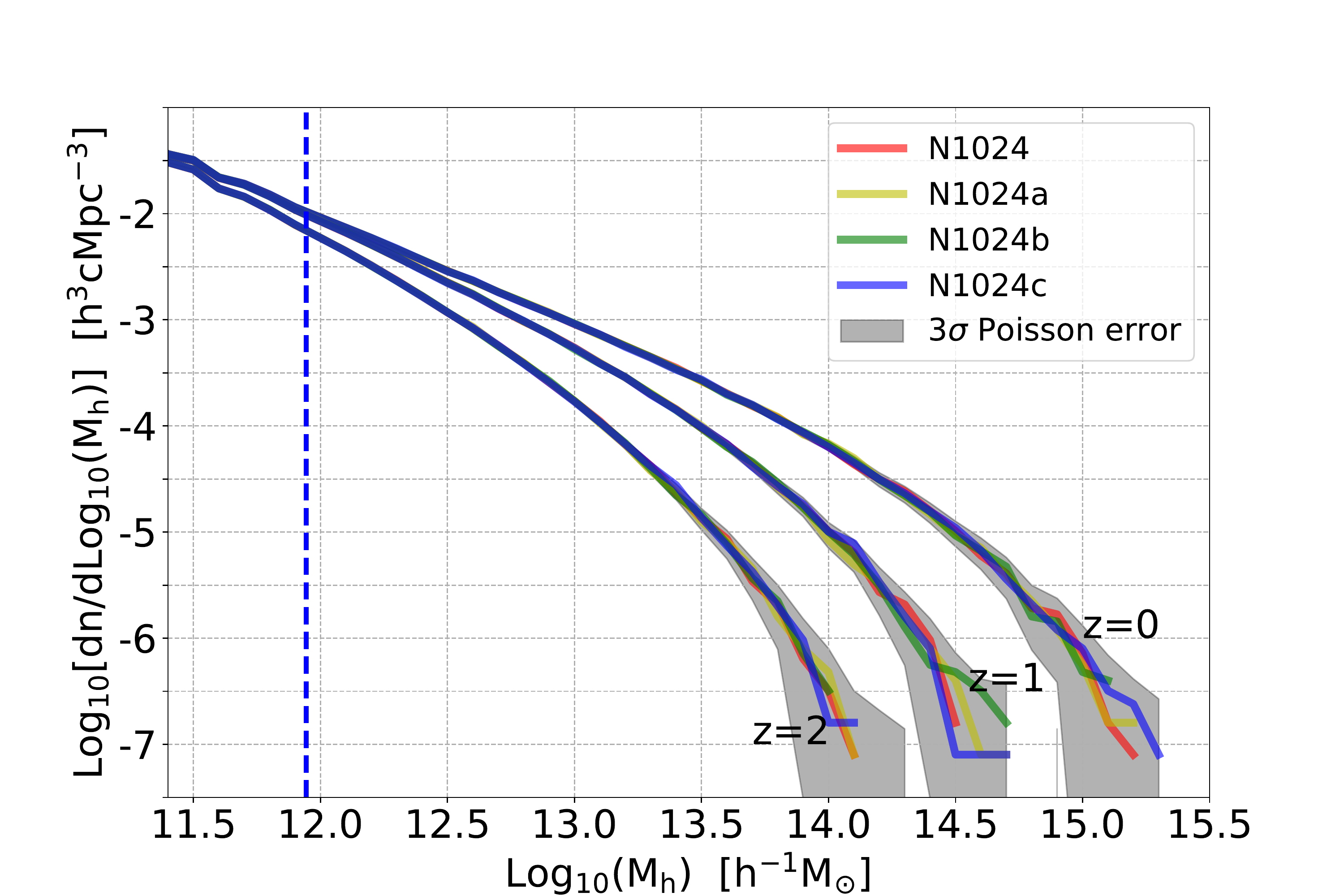}
\caption{Comparison of the halo mass function in the twin simulations. The blue dashed line denotes the halo mass with $100$ particles, and the gray shaded region marks the $3\sigma$ scatter, i.e., the range from $|N-3\sqrt{N}|$ to $N+3\sqrt{N}$, where $N$ is the mean halo count in each mass bin. At all redshifts, four curves are highly overlapped except at the high mass end.}
\label{fig:fig2}
\end{figure}

Firstly, we show the halo mass function as an important statistical test of our twin simulations. Since the amplitude of the initial power spectrum of the twin simulations is exactly the same, one would expect that the difference in their halo mass function is very small. On the other hand, each N-body simulation has a finite box size, hence the cosmic variance between different simulations will produce different mass functions. Fig. \ref{fig:fig2} shows the halo mass function of N1024, N1024a, N1024b and N1024c. As we can see, all curves are nearly identical except at the high mass end, which can be explained by the effect of cosmic variance (well within the range of grey shadows). And because of the same reason, any similar figure regarding the statistical properties of the twin simulations gives nearly the same result. So we will not distinguish them hereafter.

\begin{figure*}[htp]
\plotone{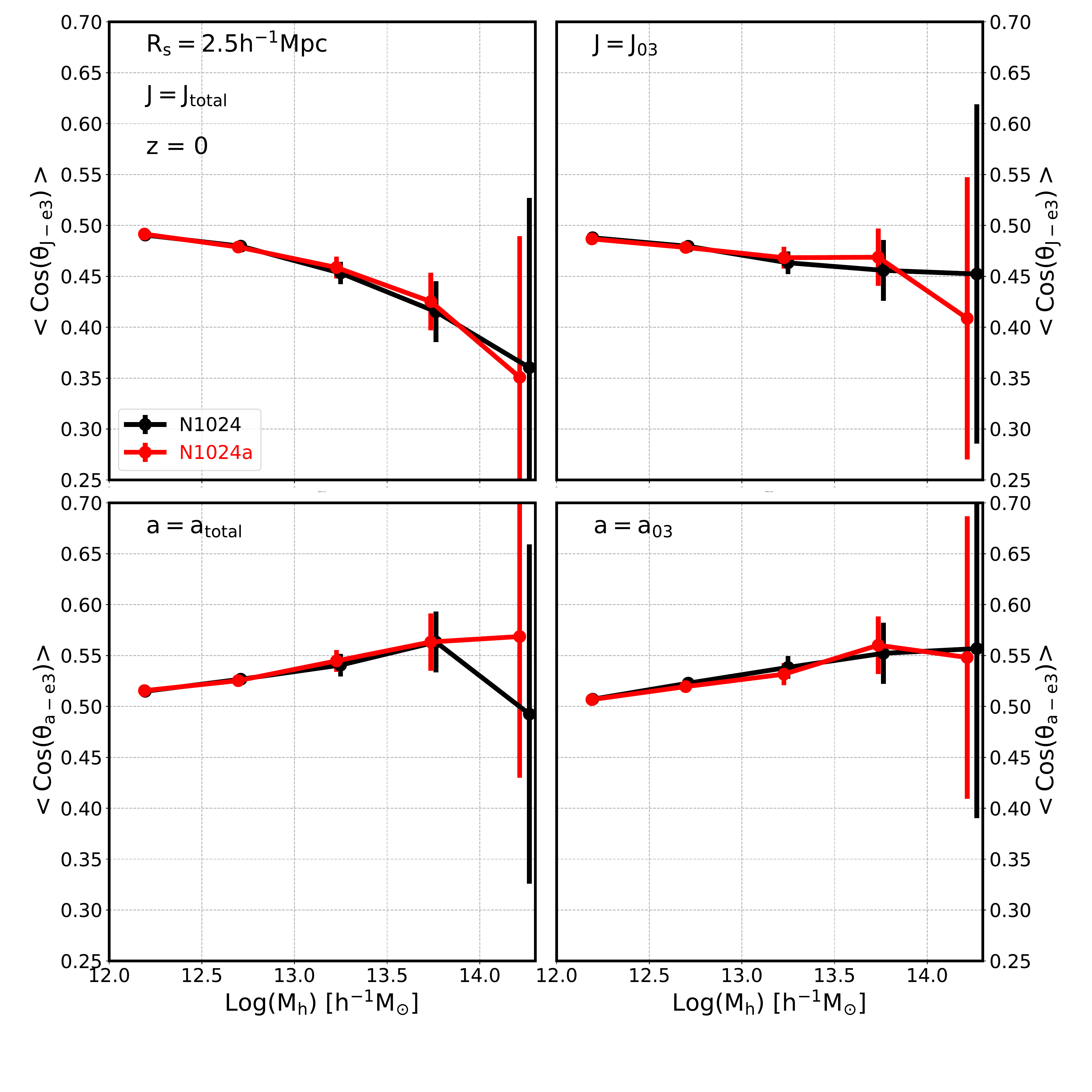}
\caption{The correlations between halo spin/major axis with the large-scale structure ($\textbf{e}_3$ vector) from the twin simulations at $z=0$. In each panel the black and red lines are from the two simulations respectively. Upper two panels show the mean cosine value of the angle between halo spin ($\rm J$ and $\rm J_{03}$, calculated at 0.3 times halo virial radius) and $\textbf{e}_3$ vector, lower two panels show the correlations between major-axis ($\rm a$ and $\rm a_{03}$) and $\textbf{e}_3$. The $1\sigma$ poisson error bars are also shown. This figure shows that the twin simulations are statistically equal.}
\label{fig:fig3}
\end{figure*}

In Fig. \ref{fig:fig3} we show another statistical test regarding the mass dependence of correlations between halo spin or halo major-axis with the $\textbf{e}_3$ vector for matching halo pairs in the twin simulations. The upper two panels are for halo spin, while the lower two panels are for halo major-axis. This figure shows that the correlations between the twin simulations agree with each other within $1\sigma$ scatter, showing again that the twin simulations are statistically equivalent. In the left panels the halo spin and major axis are calculated using all particles within the virial radius of the halo. The trend of mass dependence of spin-$\textbf{e}_3$ correlation agrees well with previous works \citep{2012MNRAS.427.3320C, 2016IAUS..308..421P, 2019ApJ...872...37L}. Note that here we only use halos with more than 100 particles ($\sim10^{12}\mathrm{M}_{\odot}/h$), roughly corresponding to the flip mass of the spin-LSS correlation found by previous works \citep[e.g.,][and references within]{2018MNRAS.473.1562W}.

In N-body simulations, the direction of halo spin is often slowly twisted along the halo radius, i.e., there is a misalignment between the inner spin and the total spin \citep[e.g.,][]{2020arXiv200503025S}. Motivated by the fact that the baryons experience the same tidal forces as the dark matter, the spin of the galaxies embedded in halos usually follows the inner spin of the host halo. To ensure a relatively reliable result, we calculate the spin and major-axis vectors at $0.3$ times the virial radius to describe the inner properties of matching halo hereafter. The right two panels of Fig. \ref{fig:fig3} show the mass dependence of inner properties (i.e. spin and major-axis at $0.3$ times the virial radius). It is seen that, comparing to the alignment of total spin and $\textbf{e}_3$ vector, the correlation of inner spin and $\textbf{e}_3$ is nearly independent of the mass. On the other hand, as mentioned by \citet{2018MNRAS.473.1562W}, the halo spin-$\textbf{e}_3$ correlation is stronger at high redshift. Considering that the inner spin is inherited from the total spin at high redshift, the alignment signal of inner spin and $\textbf{e}_3$ vector is partly lost due to the nonlinear process in the center of high mass halos at later time.

\subsection{Match ratio and mergers} \label{sec:3.2}

\begin{figure*}[htp]
\plotone{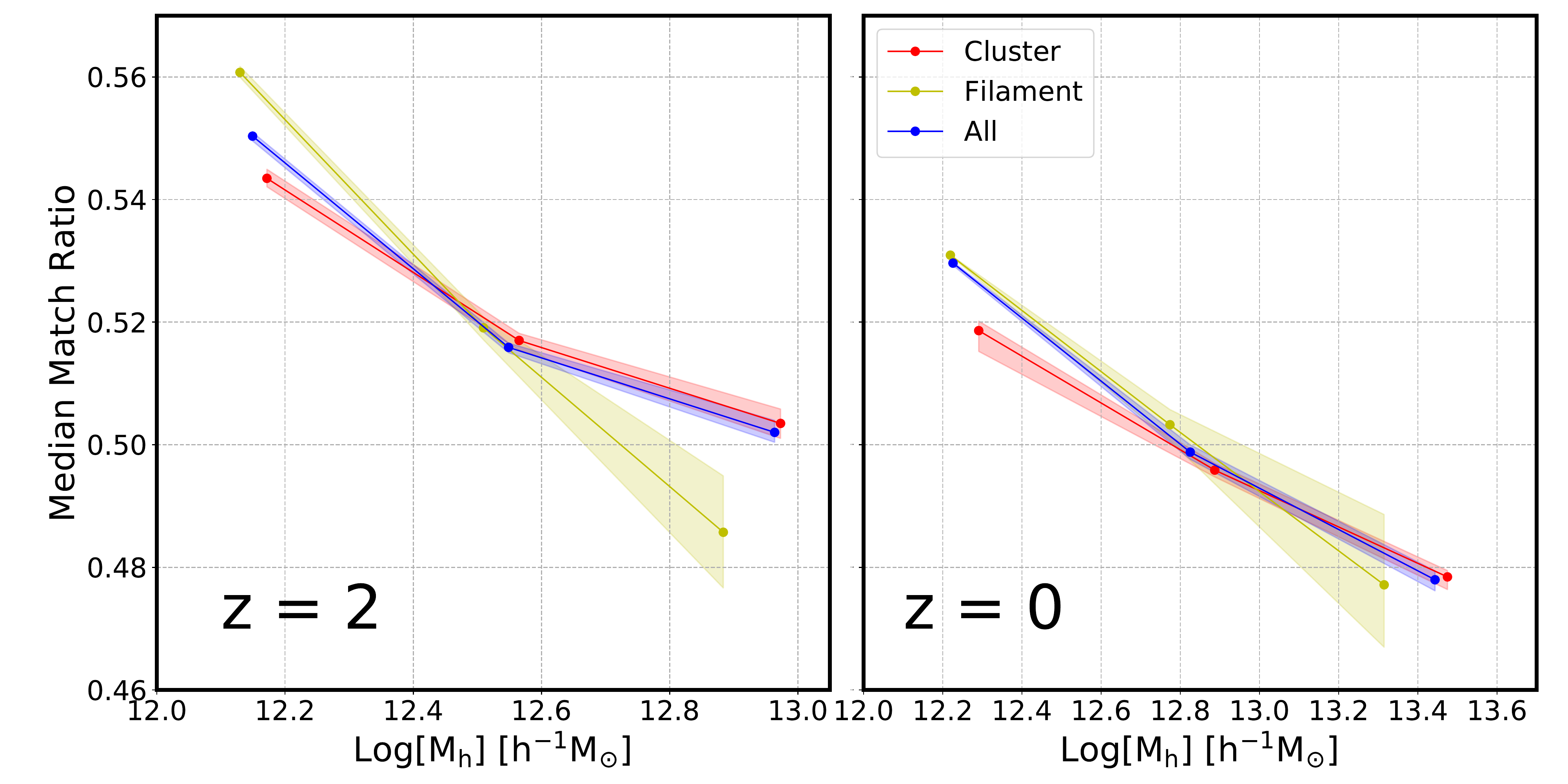}
\caption{Mass dependence of the match ratio $R$ of halos in the twin simulations. Blue lines are for all matched halo pairs. Red and yellow lines are for matching halo pair located in the same environment (here we choose cluster and filament). The shaded region marks the maximum and minimum of $6$ twin simulation pairs, while the solid line shows the mean value.}
\label{fig:fig4}
\end{figure*}

In this section we focus on the matching halo pairs between the twin simulations. Firstly in Fig. \ref{fig:fig4}, we show the mass dependence of match ratio $R$ in $6$ twin simulation pairs (i.e., matched halos between N1024 and N1024a, N1024 and N1024b, N1024 and N1024c, N1024a and N1024b, N1024a and N1024c, N1024b and N1024c) at $z=0$ and $z=2$. The shaded region marks the maximum and minimum median match ratio, which roughly illustrate the systematic errors introduced by cosmic variance. For each twin simulation pair, the mass dependence of match ratio $R$ is for the first simulation. If the mass bins are sorted by the second one, as we mentioned in Sec. \ref{sec:2}, the conclusions are still the same. The lower limit of the y-axis is set to $0.4$, which corresponds the arbitrary threshold imposed to select matching halo pairs. It can be seen that the match ratio of low mass halos is relatively higher, and comparing the same line of both panels, we also find that halos at high redshift usually have higher match ratios. For massive halos at lower redshift, the frequent mergers or mass accretion along the large-scale structure decrease the match ratio between the twin simulations. The yellow and red lines are for matching halos located in the different environments in both simulations. The median match ratio $R$ of halos in filaments decreases much faster compared to those in clusters, indicating that halos in filaments are more influenced by mass accretion (along the filament) than in clusters.

Here we shortly explain how the value of match ratio is reduced in minor merger events. Suppose that halo $A$ merges with another halo $b$ in N1024, while the matching halo $A'$ and $b'$ in N1024a are not involved in any merger event. At the end of the minor merger, the dark matter particles of $A$ and $b$ mix together. The information about the matching halo pair $b$ and $b'$ in the later snapshots is completely lost, and the match ratio of halo $A$ and $A'$ will be reduced because of the external particles from halo $b$.

\subsection{The evolution of spin with LSS} \label{sec:3.3}

\begin{figure}[htp]
\plotone{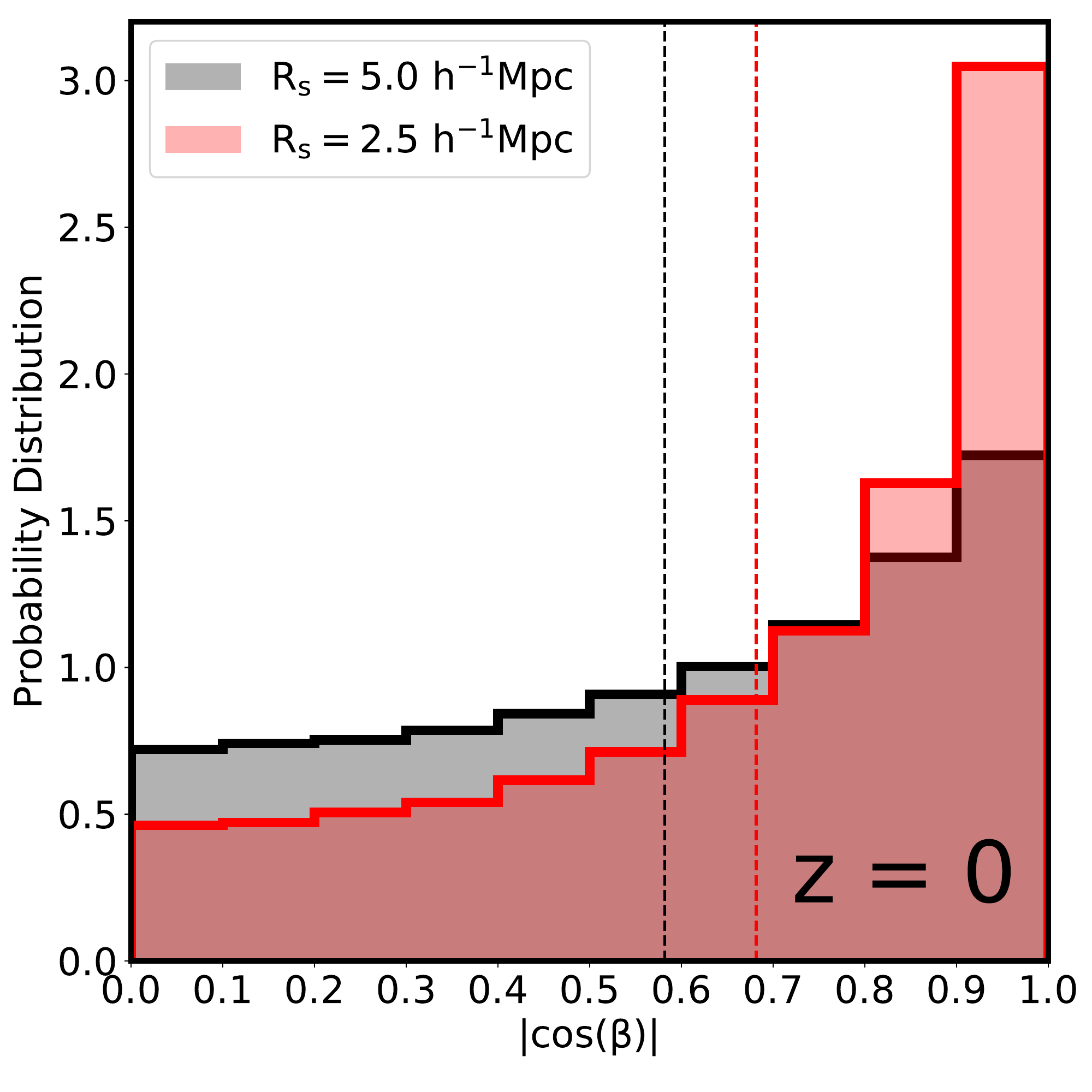}
\caption{The probability distribution of the rotation angle, $\beta$, between the large-scale environment ($\textbf{e}_3$ vector) of matching halo pairs in N1024 and N1024a. Here results with two different smoothing lengths are shown, and the vertical dashed lines denote the mean value of corresponding distributions.}
\label{fig:fig5}
\end{figure}

\begin{figure*}[htp]
\plotone{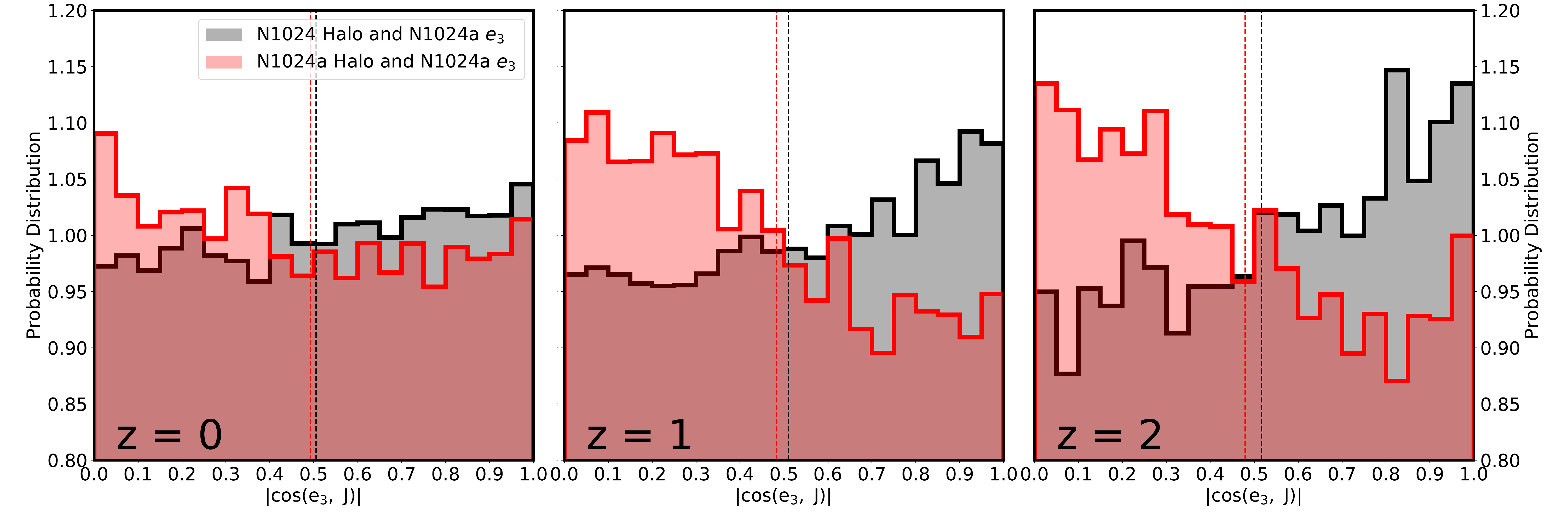}
\caption{The probability distribution of absolute cosine value of the angle between halo spin and $\textbf{e}_3$ vector in the twin simulations at different redshifts. For the red histograms, both halo spin and $\textbf{e}_{3}$ are from the same simulation (N1024a), and the black histograms show the cross correlation between halo spin from N1024 and the $\textbf{e}_{3}$ vector from N1024a. If halo spin does not respond to change of LSS, the spin-LSS correlation can not be maintained.}
\label{fig:fig6}
\end{figure*}

As we mentioned in Sec. \ref{sec:2.1}, the $\textbf{e}_{3}$ obtained from the hessian matrix indicates the direction of the LSS around a halo. For matching halo pairs in the twin simulation, we calculate the rotational angle of the corresponding eigenvectors between the twin simulation as,
\begin{equation}
\label{equ:equ6}
\cos\left(\beta\right)=\frac{\textbf{e}_3\cdot \textbf{e}_3'}{\vert\textbf{e}_3\vert\vert\textbf{e}_3'\vert}\;,
\end{equation}
here $\textbf{e}_3$ and $\textbf{e}_3'$ are the corresponding eigenvectors of matching halo pairs. In Fig. \ref{fig:fig5} the red histogram shows the distribution of the rotation angle between the $\textbf{e}_{3}$ vector in the twin simulation pair, N1024 and N1024a. It is found that there is more probability to have halo pairs for which the rotation angle of $\textbf{e}_{3}$ is not significant, which suggests that although the power spectrum on large scales ($>20$ Mpc/h) is totally different in the twin simulations, the nearby local environment smoothed on 2.5 Mpc/h is more or less maintained by the matched halo pairs. The black histogram shows a comparison with a larger smoothing length  with $R_s=5.0$ Mpc/h. In this case the rotation angle of large-scale environment becomes larger, but still maintaining some level of parallel signal.

In this work we are mainly interested in how halo properties, spin or shape, correspond to the change of the large-scale environment, in particular, the direction of the LSS given by $\textbf{e}_{3}$. To clearly select matching halo pairs with significant change of their large-scale environments, we select halo pairs from the twin simulations that have:
\begin{equation}
\label{equ:equ7}
\vert\cos\left(\beta\right)\vert\leq 0.707\quad\mathrm{or}\quad \beta\geq45^{\circ}\;.
\end{equation}
In this way we are looking at halo pairs with significant rotation of their $\textbf{e}_{3}$ vectors. Of course, we can select halo pairs with larger rotation angles to increase the signal, but that will reduce the sample size.

Based on Fig. \ref{fig:fig3}, halos in twin simulations show the same spin-LSS correlation. We can then ask what will happen if the original spin (of halo in, for example, N1024) does not change, can it still maintain a similar correlation with the new $\textbf{e}_3$ from other twin simulation? If not, it means that the spin needs to be changed systematically to match the correlations (as shown in Fig. \ref{fig:fig3}). This can prove that the correlation between spin and LSS is mainly affected by large-scale structures. 

In Fig. \ref{fig:fig6} we plot the distribution of the spin-LSS correlation at three different redshifts for halo pairs with larger rotation angles given by Eq. \ref{equ:equ7}. The red shaded areas are the PDFs of halos whose spin and $\textbf{e}_3$ are from the same simulation N1024a. The grey histograms are the cross correlations of the halo spin-$\textbf{e}_3$ correlation, where halo spin is from N1024, but the $\textbf{e}_3'$ vector is from N1024a. For matching halo pairs, the spin-LSS correlation in the same simulation, N1024a, is similar to what is shown in Fig. \ref{fig:fig3} and this is not surprising as the selected halos are a subsample of those shown in Fig. \ref{fig:fig3}. It is interesting to note that the cross correlations (black histograms) are reversed, showing a different trend with the red histograms. This result shows that the rotation of $\textbf{e}_3$ vectors will lead to proper rotation of the halo spin so to maintain the original halo spin-$\textbf{e}_3$ correlation seen in one simulation. It indicates that the halo spin is mainly affected by the orbital angular momentum of accreted mass from the LSS along the $\textbf{e}_3$ vector.

\subsection{Matching halos and $\textbf{e}_3$ rotation angle} \label{sec:3.4}


\begin{figure*}[htp]
\plotone{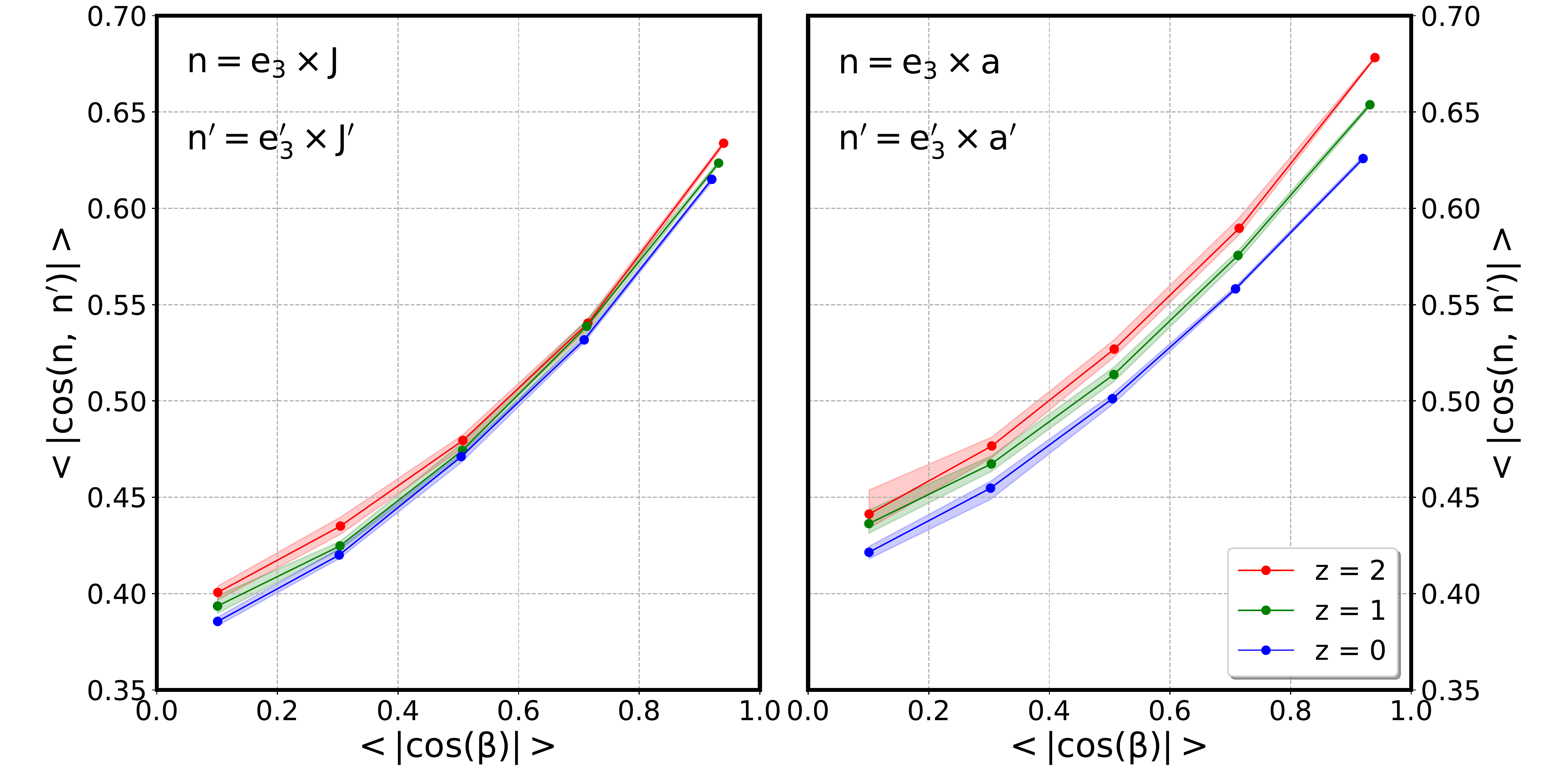}
\caption{The relationship between the rotation of normal vector of the spin-$\textbf{e}_3$ system and the rotation of LSS at different redshifts. The rotation of LSS does have effect on the spin/major-axis of halo. Shaded regions and solid lines are plotted in the same way as in Fig. \ref{fig:fig4}. See text for more detail.}
\label{fig:fig8}
\end{figure*}

From Sec. \ref{sec:3.3} we learn that the spin/major-axis will adjust its direction appropriately when the $\textbf{e}_3$ vector changes across twin simulations. In this subsection we will explore this reaction in more detail. To maintain the spin-LSS correlation, one may intuitively expect that the rotation angle of halo spin/major-axis will depend strongly on the rotation angle $\beta$ (see Sec. \ref{sec:3.3}, Eq. \ref{equ:equ6}). More specifically, we expect that the rotation angle of spin is larger for halos with larger $\beta$ so as to maintain the spin-$\textbf{e}_{3}$ relation. However, if we check the correlation mentioned above, we will find that the rotation of spin/major-axis is nearly independent of angle $\beta$. This is a bit surprising and it is related to the 3D space configuration of halo spin and LSS. Even in 2D space, it is easy to configure a case where the $\textbf{e}_3$ vector changes significantly while the halo spin/major-axis does not need any adjustment to maintain the same spin-LSS correlation. The situation is more complicated in the 3D case. To better show how this could happen, we introduce the cross-product between $\textbf{e}_3$ vector and spin/major-axis as the normal vector of halo-LSS system:
\begin{equation}
\label{equ:equ8}
\textbf{n}=\textbf{e}_3\times\textbf{a}\;,
\end{equation}
where $\textbf{a}$ is the halo spin or major-axis vector. Under this definition, the rotation of the $\textbf{e}_3$ vector will lead to the corresponding rotation of the \textbf{n} vector in most cases. In Fig. \ref{fig:fig8} we show the relationship between the rotation angle of $\textbf{e}_{3}$ and the normal vector defined above. Here we can see that a strong positive proportional relationship between the rotation angle $\beta$ and \textbf{n} vectors for all redshifts, i.e., the change of $\textbf{e}_3$ vector will lead to corresponding rotation of the whole halo-LSS system, in which the correlation between spin/major-axis and $\textbf{e}_3$ vector always holds. This gives us a picture of how halo spin/major-axis reacts according to the change of LSS. On the other hand, the proportional relationship in both two panels are nearly independent of the redshift (only the \textbf{n} vector from major-axis shown in the right panel has a weak evolution from $z=2$ to the present-day). This implies an  universality for the picture described above.

\subsection{Halo shape parameters with LSS} \label{sec:3.5}

\begin{figure*}[htp]
\plotone{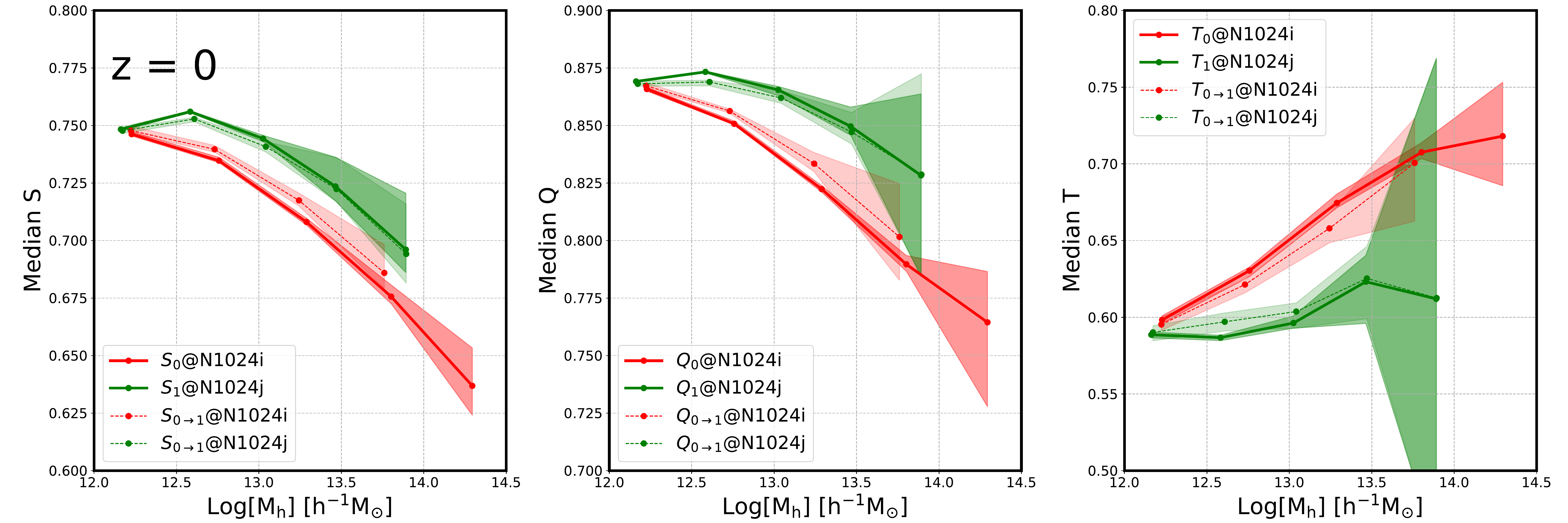}
\caption{Mass dependence of halo shape parameters (S, Q, T) at $z=0$. The subscript denotes the LSS environment of matching halo pairs, i.e., $0$ means cluster, $1$ means filament, and $0\rightarrow1$ stands for the environment transformation from cluster to filament for the matching halo pair. i and j denote the twinned simulations in each twin simulation pair. All the shaded regions, solid lines and sorted mass bins here are plotted in the same way as in Fig. \ref{fig:fig4}. The halo shape is mainly determined by its current large-scale environment.}
\label{fig:fig9}
\end{figure*}

Some previous works \citep[e.g.,][]{2007MNRAS.375..489H, 2020MNRAS.495..502M} have revealed the correlations between halo shape and the LSS environment. In this work we also explore the impact of LSS environment on ellipticity and triaxiality of halos ($M_h>10^{12}h^{-1}\mathrm{M}_{\odot}$). In Fig.~\ref{fig:fig9} the mass dependence of three shape parameters $S$, $Q$ and $T$ at $z=0$ are shown in three panels, respectively. We note that the results are nearly independent of redshift, therefore, we only show the $z=0$ results. The red and green lines are for halos in different twin simulations. The solid lines are for halos in specific LSS environment ($0$ stands for cluster and $1$ stands for filament). In addition, we also choose the matching halo pairs which go through an environment transformation from cluster to filament. These halos are denoted by the subscript $0\rightarrow1$. Firstly it can be seen that, statistically, the shapes of halos in different LSS environment have some systematic differences, i.e., compared to cluster halos, halos in filament ($M_h>10^{12}h^{-1}\mathrm{M}_{\odot}$) are much more spherical and more oblate, which agrees with previous results \citep[e.g.,][]{2015ApJ...813....6K}. On the other hand, lines in the same color are quite consistent for all shape parameters (reminding that there is a symmetry between twin simulations in terms of individual measurements), which proves that the current LSS environment really has dominant impact on the ellipticity and triaxiality of halos, regardless of their environments in the other symmetric simulation. Combined with results presented in the previous section on the spin-LSS correlation, our results suggest that current environment dominates the correlation between halo spin/major axis with the LSS.


\section{Conclusions and Discussion} \label{sec:4}

In this paper, we investigate the effect of LSS on halo pairs matched between twin N-body simulations. The only difference between the twin simulations is that in one run, the random seeds for the phase of small $k$ where replaced when the initial conditions are generated. Once the sample of matching halos at different redshifts is established, one can use them to separate the entangled factors of large  and small scales in the formation of dark matter halos. In this work we mainly focus on the correlations between the halo properties (spin, major-axis, shape parameters) and the $\textbf{e}_3$ vector of the Hessian matrix, which is thought to be an universal definition of the LSS. Our main results are summarized as follows,
\begin{itemize}
\item A few statistical tests, including the halo mass function and the halo spin-$\textbf{e}_3$ correlation, show that the twin simulations are statistically equal, ensuring that our technique for producing controlled twin simulation is robust and reliable.
\item The ratio of matching halo pairs in the twin N-body simulations, based on a merit function, decreases with increasing halo mass. This is due to the later stronger evolution for massive halo by merger or mass accretion along the large-scale structure which is different in the twin simulations.
\item The halo spin/major-axis will adjust its direction appropriately when LSS environment changes across twin simulations, i.e., the change of $\textbf{e}_3$ vector will lead to corresponding rotation of the whole halo-LSS system, to ensure the correlation between spin/major-axis and $\textbf{e}_3$ vector identical in the twin simulations.
\item Statistically, halos in filament ($M_h>10^{12}h^{-1}\mathrm{M}_{\odot}$) are much more spherical and more oblate than cluster halos, and the LSS environment really has dominant impact on ellipticity and triaxiality of halos.
\end{itemize}

Since our technique of running twin simulations and establishing the matching halo pairs between them are verified to be robust and reliable, in the future, this technique can be implemented in some larger N-body simulations, or even hydrodynamical simulations, to identify matching galaxies embedded in the center of matching halo pairs. Moreover, multi-scale initial conditions for cosmological simulations, e.g. the zoom-in techniques, can also be employed to study how LSS environment influences galaxy formation and evolution in more  details. On the other hand, one can degenerate the entangled factors on different scales in the halo and galaxy formation by choosing different $L_0$ in Eq. \ref{equ:equ2}. We conclude that twin simulations with controlled initial condition on different scales could be useful to entangle the formation of halo/galaxy and the large-scale environment.

\acknowledgments

We thank the anonymous reviewer for constructive and insightful suggestions that improved this paper. We acknowledge support from the joint Sino-German DFG research Project ``The Cosmic Web and its impact on galaxy formation and alignment" (NSFC No.11861131006, DFG-LI 2015/5-1 )， and the funds of cosmology simulation database (CSD) in the National Basic Science Data Center (NBSDC). HZC and XK acknowledge support from the the NSFC (No. 11825303, 11333008), the 973 program (No. 2015CB857003), and the China Manned Space Program (No. CMS-CSST-2021-A03, CMS-CSST-2021-B01). NIL acknowledges financial support of the Project IDEXLYON at the University of Lyon under the Investments for the Future Program (ANR-16-IDEX-0005).


\end{CJK*}
\end{document}